\documentclass{PoS}
\title{The Observation Campaign of SS 433 in April 2006}

\ShortTitle{Observation Campaign of SS 433 in April 2006}

\author{\speaker{T.~Kotani}$^a$, 
K.~Kubota$^b$, M.~Namiki$^c$, N.~Kawai$^a$, Y.~Ueda$^b$, 
S.~Trushkin$^d$, S.~Fabrika$^d$, V.~Afanasiev$^d$, P.~Abolmasov$^d$, 
K.~Kinugasa$^e$, T.~Nagata$^b$, T.~Irsmambetova$^f$, T.~Tsukagoshi$^g$, K.~Nakanishi$^h$, M.~Tsuboi$^h$,
S.~Ozaki$^{bi}$, K.~Yanagisawa$^j$, S.~Nishiyama$^k$, 
T.~Shimokawabe$^a$, Y.~Yatsu$^a$, T.~Ishimura$^a$, and K.~Fujisawa$^l$\\
\llap{$^a$}        Tokyo Tech, 2-12-1 O-okayama, Meguro, Tokyo 152-8551, Japan\\
\llap{$^b$}        Kyoto University, Kitashirakawa-Oiwake-cho, Sakyo-ku, Kyoto 606-8502, Japan\\
\llap{$^c$}        Osaka University, 1-1 Machikaneyama, Toyonaka, Osaka
560-0043, Japan\\
\llap{$^d$} Special Astrophysical Observatory, Nizhnij Arkhyz, Karachaevo-Cherkassia 369167, Russia\\
\llap{$^e$} Gunma Astronomical Observatory, 6860-86 Nakayama, Agatsuma, Gunma 377-0702, Japan\\
\llap{$^f$} Crimean Astrophysical Observatory, Nauchny, Crimea 98409, Ukraine\\
\llap{$^g$} SOKENDAI,
Minamimaki, Minamisaku, Nagano 384-1305, Japan\\
\llap{$^h$}  Nobeyama Radio Observatory, Minamimaki, Minamisaku, Nagano 384-1305, Japan\\
\llap{$^i$}  Nishi-Harima Astronomical Observatory, Sayo-cho, Hyogo 679-5313, Japan\\
\llap{$^j$}  Okayama Astrophysical Observatory/NAOJ/NINS, 3037-5 Honjo,  Okayama 719-0232, Japan\\
\llap{$^k$} Osaka Kyoiku University, 4-698-1
Asahigaoka, Kaibara, Osaka 582-8582, Japan\\
\llap{$^l$} Yamaguchi University, 1677-1 Yoshida, Yamaguchi, Yamaguchi 753-8512, Japan\\
E-mail: \email{kotani@hp.phys.titech.ac.jp}}

\abstract{A radio-IR-optical-X-ray observation campaign of SS 433 has
been performed in April 2006, when the jet axis is almost perpendicular
to the line of sight.  Five flares have been detected during the
campaign by radio monitoring observation with RATAN-600.  The X-ray astronomical
satellite Suzaku observed the source in and out of eclipse.  In the
X-ray data out of eclipse, the flux shows a significant variation with a
time scale of hours.  The source seems to be in the active state during
the campaign.  The observation logs and preliminary results are presented.}

\FullConference{VI Microquasar Workshop: Microquasars and Beyond\\
		September 18-22 2006\\
		Societ\`a del Casino, Como, Italy}

\begin{document}

\section{Introduction}
SS~433 is the unique microquasar known for the very stable continuous jet
emanating at a quarter of the speed of light~\cite{margon84,fabrika04}.  The optical and X-ray
spectra are abundant in pairs of Doppler-shifted emission lines from the
bipolar jets.  The emission lines are evidence that the jet plasma
contains baryons, while other microquasars' jets do not show such
Doppler-shifted emission lines, presumably because they consist of pair
plasma.

A multi-wavelength observation campaign is desirable for the comprehension
of SS~433, because the behavior and relation of the system components, a
synchrotron jet, an optically-thick accretion disk, and a high-energy jet
engine, can be studied only with a multi-wavelength campaign~\cite{kotani99,chakrabarti05,cherepashchuk05,kotani06}.
Especially, radio monitoring to diagnose the state of the system,
optical spectroscopy to determine the precessional phase, and
X-ray observation of the core of the system are essential.
We present preliminary results from an intensive multi-wavelength
observation campaign of the source in April 2006 with the X-ray observatory
Suzaku and several large radio and optical telescopes.

\section{Observations}
The source has been observed with the X-ray astronomical satellite
Suzaku~\cite{mitsuda06} at MJD = 53830 and 53833
(Table~\ref{tbl:xobslog}).  The orbital phases at the observations are
1.0 and 0.22, respectively~\cite{gladyshev87}.  An optical-IR-radio
observation campaign was performed to cover the periods of the Suzaku
observations (Table~\ref{tbl:specobslog}--\ref{tbl:radiobslog}).
Spectra have been taken with the 6-m Telescope (BTA) at the Special
Astrophysical Observatory RAS (SAO RAS), the 122-cm Telescope at the
Padova-Asiago Observatory, the 150-cm Telescope at the Gunma Astronomical
Observatory~\cite{obayashi04}, and the Nayuta Telescope at the Nishi-Harima
Astronomical Observatory.  Photometric data have been  obtained with the KGB-38
Telescope at the Crimean Astrophysical Observatory, a 40-cm Telescope at
the Kyoto University, MITSuME Akeno 50 cm~\cite{kotani05} at the Akeno
Observatory, MITSuME OAO 50 cm~\cite{kotani05} at the Okayama
Astrophysical Observatory (OAO), the 51-cm Telescope at the Osaka Kyoiku
University~\cite{yokoo94}, and telescopes in VSNET~\cite{kato04} and in the
Variable Star Observers League in Japan (VSOLJ).  Infrared photometry
data have been obtained with the IRSF 1.4-m Telescope~\cite{kandori06} at the
South African Astronomical Observatory (SAAO).  The radio activity from
1.0 GHz to 21.7 GHz has been  monitored with the RATAN-600 at SAO in the
period covering the campaign.  The 32-m radio telescope (RTF-32) at the
Institute of Applied Astronomy RAS (IAA RAS)
and the Nobeyama Millimeter Array at the Nobeyama Radio Observatory have
also
participated in the campaign.

\section{Precessional Phase}
\begin{figure}[h]
\centering
\begin{minipage}{.5\textwidth}
 \includegraphics[width=1\textwidth,angle=270]{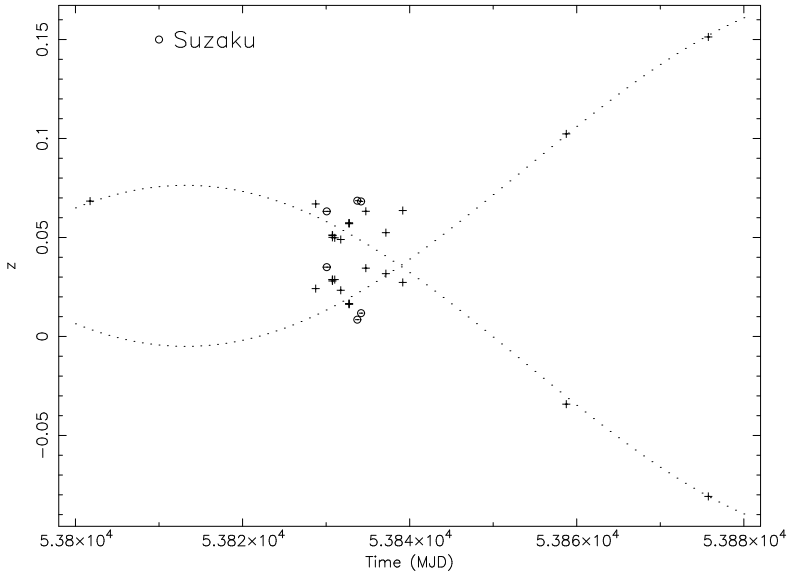}
\end{minipage}~~~~~~~~~~~~~~~~~~~~~~~~~~~~~~~~~~
\begin{minipage}{.3\textwidth}
\caption{Doppler shifts.  Optical data (crosses) and X-ray
 data (open circles) are plotted.  Dotted lines are sinusoidal curves~\cite{margon89}
 shifted to fit the observational points.}\label{fig:z}
\end{minipage}
\end{figure}

The Doppler shifts of the jets of SS~433 measured with the X-ray Imaging
Spectrometers (XIS)~\cite{koyama06} on Suzaku and the optical
spectrometers are shown in Fig.~\ref{fig:z}.  This plot provides 
information on the geometrical configuration of the system at that time.
Observed Doppler shifts
deviate from the sinusoidal curves of the five-parameter kinematic
model~\cite{margon89} due to the nodding motion~\cite{gies02}.  From
Fig.~\ref{fig:z}, it is confirmed that the campaign has been performed at a
precessional phase when the inclination angle of the jet axis slightly
exceeds 90$^\circ$.

\section{Light Curves}
\begin{figure}[h]
\centering
\begin{minipage}{.5\textwidth}
 \includegraphics[width=1\textwidth,angle=270]{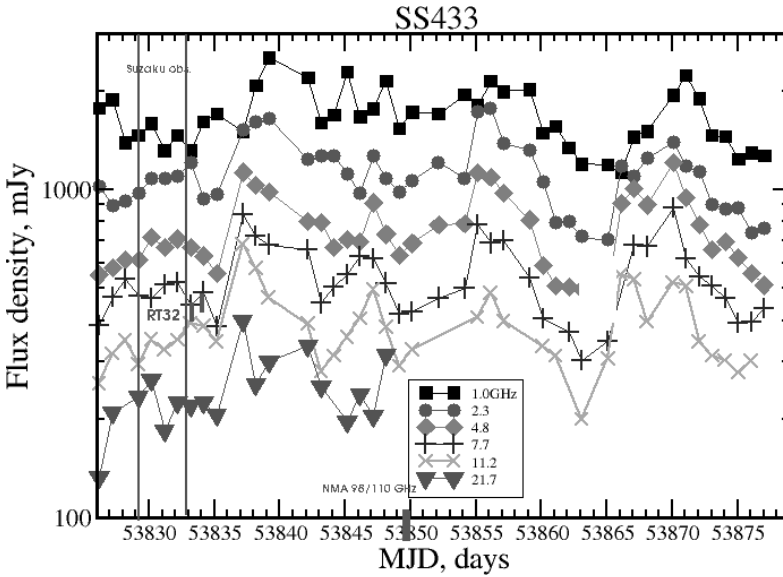}
\end{minipage}~~~~~~~~~~~~~~~~~~~~~~~~~~~~~~~~~~
\begin{minipage}{.3\textwidth}
\caption{Radio light curves obtained with RATAN-600/SAO\@.  The data
obtained with RTF-32/IAA (MJD = 53833 and 53834) and NMA/Nobeyama (MJD =
53850) are also plotted.  The epochs of Suzaku's observations are
indicated by vertical lines.}\label{fig:radiolc}
\end{minipage}
\end{figure}

Radio monitoring of the activity of the source is essential to a
multi-wavelength campaign~\cite{trushkin04}.
The radio light curves during the campaign are shown in Fig.~\ref{fig:radiolc}.  The curves
exhibit five radio flares at MJD = 53837, 53847, 53855, 53866, and
53870, suggesting that the source has been  in the active state successively
ejecting massive jets.  Optical spectroscopic observations with the 122-cm Telescope/Padova-Asiago and
Nayuta/Nishi-Harima coincide with the flare at MJD = 53837.  It is
interesting that no lines are detected from the coinciding observation
of Nayuta.  The observations with Suzaku have been performed before the
first detected flare.  At that time, the 2 GHz flux densities fluctuate
around 1 Jy, indicating a high activity even before the first detected flare.  Suzaku might have
observed the source in the interval between two flares.

\begin{figure}[h]
\centering
\begin{minipage}{.3\textwidth}
 \includegraphics[width=1\textwidth,angle=270]{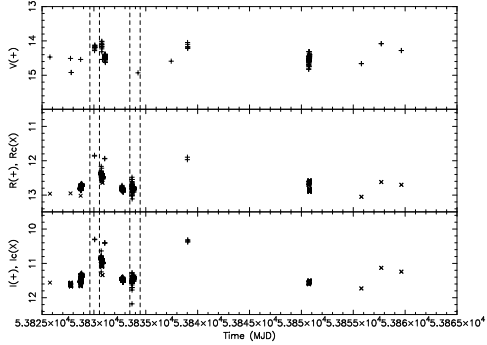}
\end{minipage}~~~~~~~~~~~~~~~~~~~~~~~~~
\begin{minipage}{.45\textwidth}
\caption{Optical light curves.  The periods of two Suzaku observations are
indicated by vertical dashed lines.}\label{fig:optlc}
\end{minipage}
\end{figure}

The VRI optical light curves are shown in Fig.~\ref{fig:optlc}. The
participating observatories/organizations are listed in
Table~\ref{tbl:photobslog}.  A continuous observation with  MITSuME
OAO 50 cm coincides with the Suzaku observation out of eclipse.

\begin{figure}[h]
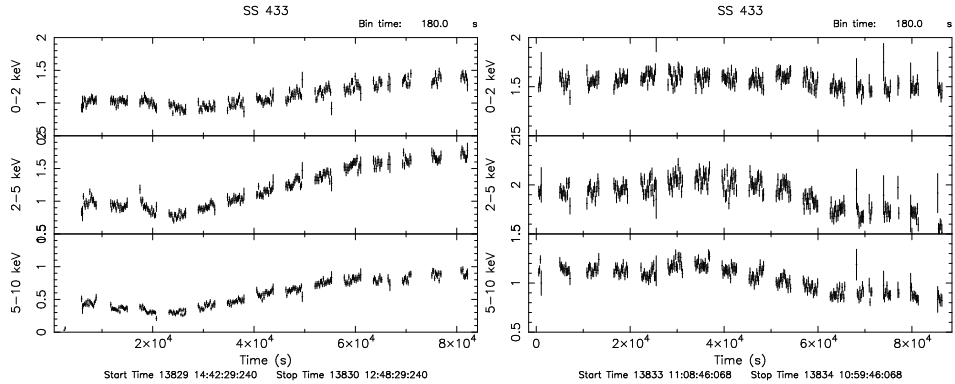

\centering
\includegraphics[width=.33\textwidth,angle=270]{in_src3_lc.ps}
 \includegraphics[width=.33\textwidth,angle=270]{out_src3_lc.ps}
\caption{X-ray light curves obtained with XIS/Suzaku.  Left: 2006/04/04 (MJD = 
53829). Right: 2006/04/08 (MJD = 53833).}\label{fig:xraylc}
\end{figure}

The X-ray light curves are shown in
Fig.~\ref{fig:xraylc}. The first observation has been performed during an
eclipse.  The X-ray flux recovers from the minimum, which is slightly
different from the prediction by \cite{gladyshev87}.  The cause of the
shift   is not known yet.  In the second
observation, the flux shows a significant variability, especially in the
hard band above 5 keV\@.

\begin{figure}[h]
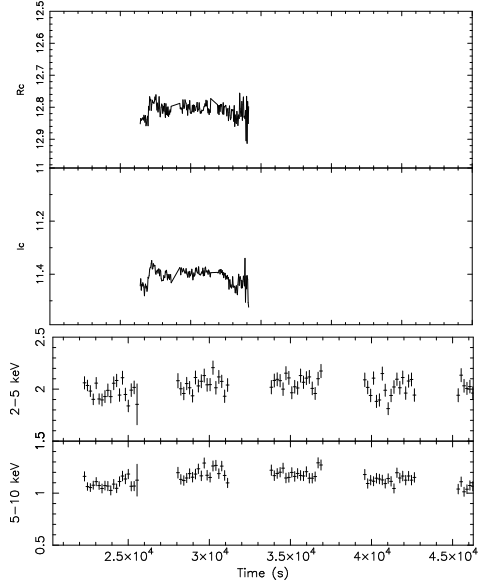

\centering
\begin{minipage}{.3\textwidth}
\begin{flushright}
 \makebox[1.4\textwidth][r]{\includegraphics[width=.95\textwidth,angle=270]{lc.mjd-53000.ps}}
 \makebox[1.415\textwidth][r]{\includegraphics[width=.74\textwidth,angle=270]{out_src3_lc.mjd833.7_834.ps}}
\end{flushright}
\end{minipage}~~~~~~~~~~~~~~~~~~~~~~~~~
\begin{minipage}{.45\textwidth}
\caption{Simultaneous X-ray and optical observations.  From top to
 bottom: Rc magnitude obtained with MITSuME OAO 50 cm, Ic with MITSuME
 OAO 50 cm, 2--5 keV count rate obtained with XIS/Suzaku, and 5--10 keV
 count rate with XIS/Suzaku. The horizontal range is from MJD = 53833.7 to 53834.0.}\label{fig:xrayoptlc}
 
\end{minipage}
\end{figure}

An example of simultaneous X-ray/optical observation is shown in
Fig.~\ref{fig:xrayoptlc}.  Correlation between these bands is to be
searched.  The optical emission from a geometrically thick
super-critical accretion disk may precede  to the X-ray emission
from the hot base of the jet, while the emission from the downstream
optical jet will lag behind the X-ray.

\section{Radio spectra}
\begin{figure}[h]
\centering
\begin{minipage}{.3\textwidth}
 \includegraphics[width=1\textwidth,angle=270]{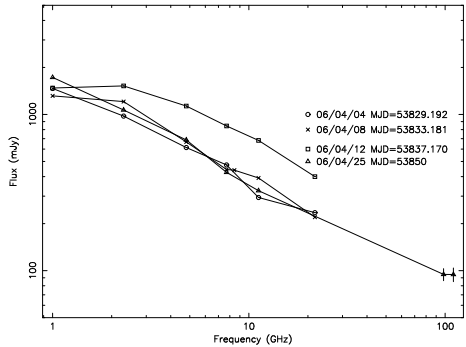}
\end{minipage}~~~~~~~~~~~~~~~~~~~~~~~~~
\begin{minipage}{.45\textwidth}
\caption{Radio spectra sampled at MJD = 53829.192, 53833.181, 53837.170,
 and 53850.}\label{fig:radiospec}
\end{minipage}
\end{figure}

Several radio spectra are plotted in Fig.~\ref{fig:radiospec}.  The
sampled dates are MJD = 53829.192 (coinciding with Suzaku's observation
in eclipse), 53833.181 (coinciding with Suzaku's observation out of
eclipse), 53837.170 (a flare peak), and  53850 (with NMA data up to
110.21 GHz).  The spectrum at the flare peak is flatter, which is a
characteristics of optically thick synchrotron sources such as small
expanding jet plasma.  Other spectra are consistent with that of optically thin
synchrotron emission with a spectral index of $\sim 0.6$.  It should be
noted that the power-law like spectrum is extended up to 110.21 GHz.
And even the flux densities other than the flare peak are higher than
previously observed values (see, e.g., Fig.~3
in~\cite{seaquist82}).  It also supports the assumption that the
source has been in the active state during the campaign.

\section{Multi-wavelength Spectrum}
\begin{figure}[h]
\centering
\begin{minipage}{.3\textwidth}
 \includegraphics[width=1\textwidth,angle=270]{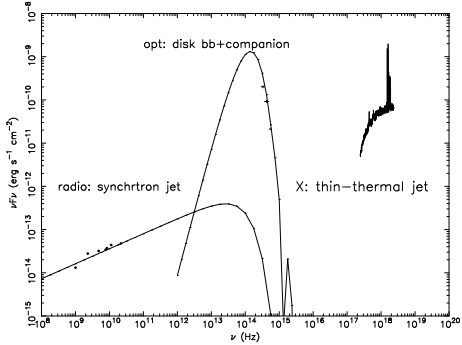}
\end{minipage}~~~~~~~~~~~~~~~~~~~~~~~~~
\begin{minipage}{.45\textwidth}
\caption{Exactly simultaneous multi-wavelength spectrum taken at MJD =
 53833.  Radio:  RATAN-600 and RTF-32.  Optical:  VSNET and MITSuME OAO 50
 cm.  X-ray:  XIS/Suzaku.  A spectral model for each energy band is plotted in solid line.}\label{fig:multispec}
\end{minipage}
\end{figure}

An exactly simultaneous multi-wavelength spectrum taken at MJD = 53833
is plotted in Fig.~\ref{fig:multispec}.  Hard X-ray data of HXD/Suzaku
and IR data of IRSF are not yet reducted.  A spectral model
for each energy band is plotted: The radio flux densities are expressed
with a power-law model attenuated by interstellar matter of $A_{\rm
v}=8$ or $N_{\rm H}= 1.57\times 10^{22}$ cm$^{-2}$.  The optical model
is the sum of a multicolor disk model with $T_{\rm in} = 10^5$ K and a
blackbody emission from a companion star with $T = 1.5\times 10^4$ K,
both of which are attenuated by interstellar matter of $A_{\rm v} = 8$.
The X-ray model consists of  bremsstrahlung continuum and emission lines attenuated by
$N_{\rm H} = 1.3 \times 10^{23}$ cm$^{-2}$.

\section{Summary}
A radio-IR-optical-X-ray observation campaign has been performed in April
2006, when the inclination of the jet axis slightly exceeds 90$^\circ$.
SS~433 has been active in the radio band and exhibited five massive jet ejection
events during the campaign.  One of the ejection events coincides with
optical spectroscopic observations.  The source in and out of eclipse
has been observed with the
X-ray astronomical satellite Suzaku.  In the data
out of eclipse, the X-ray flux shows variation of a factor of 2 in a
day, suggesting that the source are in the active state like that
observed in 1979 with Einstein~\cite{seaquist82,band89}.  The exactly
simultaneous multi-wavelength spectrum can be expressed with a
combination of three component, a synchrotron power-law model in the
radio band, a multi-color blackbody model in the optical band, and a
thin-thermal plasma model in the X-ray band.  Temporal and
spectral investigation are in progress.

\acknowledgments
TK is supported by a 21st Century COE
        Program at Tokyo Tech ``Nanometer-Scale Quantum Physics'' by the
        Ministry of Education, Culture, Sports, Science and Technology.
        This work is supported by the
        Japan-Russia Research Cooperative Program of Japan Society for
        the  Promotion of Science.  The studies are supported by the
	Russian  Foundation  Base Research (RFBR) grant N~05-02-17556
	and the mutual RFBR and Japan Society for the Promotion of
	Science (JSPS) grant N~05-02-19710.

\begin{table}[h]
\centering
{\footnotesize
 \begin{tabular}{llllcl}
\hline
\hline
\multicolumn{2}{c}{Start} & \multicolumn{2}{c}{End} &{Expos.}& Remark\\
 &\multicolumn{1}{c}{(MJD)} &&\multicolumn{1}{c}{(MJD)}&{(ks)} \\
\hline\hline
\multicolumn{6}{l}{Observatory:  Suzaku~\cite{mitsuda06}.  PI:  N. Kawai.}\\
2006/04/04 14:40 &(53829.6108) &2006/04/05 12:47 &(53830.5326)&40 &Med Eclipse.\\
2006/04/08 11:04 &(53833.4610) &2006/04/09 10:59 &(53834.4578)&40 &\\
\hline\hline
\end{tabular}}
\caption{X-Ray observation log.}\label{tbl:xobslog}
\end{table}

\begin{table}
\centering
{\footnotesize
 \begin{tabular}{llrl}
\hline\hline
\multicolumn{2}{c}{Start} & \multicolumn{1}{c}{Exposure}& Remark\\
 &\multicolumn{1}{c}{(MJD)}&\multicolumn{1}{c}{(s)} \\
\hline\hline
\multicolumn{4}{l}{Telescope:  BTA\@.  Observatory:  SAO RAS\@.  PI:  S. Fabrika.}\\
2006/04/06 00:45:35 &(53831.0317) &590 &\\
\hline
\multicolumn{4}{l}{Telescope:  122 cm.  Observatory:  Padova-Asiago.  PI:  T. Iijima.}\\
2006/04/12 03:12:57 &(53837.1340)&1200 &\\
2006/04/14 02:37:01 &(53839.1090)&1200 &\\
\hline
\multicolumn{4}{l}{Telescope:  150 cm~\cite{obayashi04}.  Observatory:  Gunma.  PI:  K. Kinugasa.}\\
2006/03/07 18:00  &(53801.75)&120&\\
2006/04/02 18:00  &(53827.75)&180&Line not detected.\\
2006/04/05 18:00  &(53830.75)&120&\\
2006/04/06 18:00  &(53831.75)&300&\\
2006/04/07 18:00  &(53832.75)&300&\\
2006/04/09 18:00  &(53834.75)&300&\\
2006/05/03 18:00  &(53858.75)&180&\\
2006/05/20 18:00  &(53875.75)&180&\\
\hline
\multicolumn{4}{l}{Telescope:  Nayuta.  Observatory:  Nishi-Harima.  PI: S. Ozaki.}\\
2006/04/03 18:04:52 &(53828.7534)&1800  &\\
2006/04/05 17:55:00 &(53830.7465) &$1800\times2$\\
2006/04/07 17:48:03 &(53832.7417) &592, 540\\
2006/04/12 08:07:16 &(53837.3384) &1800&Line not detected.\\
\hline\hline
 \end{tabular}}
\caption{Spectroscopic observation log.}\label{tbl:specobslog}
\end{table}

\begin{table}
\centering
{\footnotesize
 \begin{tabular}{llll}
\hline\hline
\multicolumn{1}{c}{Start} & \multicolumn{1}{c}{End}&Frames &Remark\\
\multicolumn{1}{c}{(MJD)}  &\multicolumn{1}{c}{(MJD)} &\\
\hline\hline
\multicolumn{4}{l}{Instrument: KGB-38.  Observatory:  Crimean.  PI:T. Irsmambetova.}\\
53830.05325 &53830.08605 &V: 10, R: 2, I: 2 &\\
53831.03587 &53831.10694 &V: 45, R: 2, I: 2\\
53839.00955 &53839.04312 &V: 6, R: 2, I: 2\\
\hline
\multicolumn{4}{l}{Telescope:  40 cm.  Observatory: Kyoto U\@.  PI: K. Kubota.}\\
53826.72946	&53826.83986	&C: 243\\
53828.72057	&53828.79068	&C: 91\\
53841.67304	&53841.82313	&C: 316\\
53848.74419	&53848.81521	&C: 171\\
53850.79558	&53850.81689	&C: 55\\
53852.80954	&53852.82616	&C: 51\\
53886.72074	&53887.24026	&C: 51\\
\hline
\multicolumn{4}{l}{Telescope:  MITSuME Akeno 50 cm~\cite{kotani05}.  Observatory:  Akeno.  PI:  T. Shimokawabe}\\
53827.76578     &53827.80385    &V: 47, I:  47\\
53828.72676     &53828.79979    &V: 91, I:  91\\
53830.73239     &53830.80104    &V: 90, I:  90\\
\hline
\multicolumn{4}{l}{Telescope: MITSuME OAO 50 cm~\cite{kotani05}.  Observatory:  OAO\@.  PI: S. Yanagisawa.}\\
53828.68783	&53828.83571	&g': 321, Rc: 321, Ic: 321\\
53830.69849	&53830.84740	&g': 360, Rc: 360, Ic: 360\\
53832.70071	&53832.84003	&g': 340, Rc: 340, Ic: 340\\
53833.67184	&53833.84119	&g': 220, Rc: 220, Ic: 220\\
\hline
\multicolumn{4}{l}{Telescope: 51 cm~\cite{yokoo94}.  Observatory: Osaka Kyoiku U\@.  PI: S. Nishiyama.}\\
53825.78	&53825.80	&V: 1, R: 1, I: 1 &\\
53828.77	&53828.79	&V: 1, R: 1, I: 1\\
53849.73734	&53849.77499	&V: 1, R: 15, I: 1\\
53850.64105	&53850.76927	&V: 1, R: 30, I: 2\\
53855.72479	&53855.75851	&R: 38\\
53857.76686	&53857.78237	&R: 14\\
53858.68045	&53858.68667	&R: 10\\
53859.69379	&53859.20651	&R: 19\\
\hline
\multicolumn{4}{l}{Organization:  VSNET~\cite{kato04}.}\\
53823.79792	&53823.80069	&V: 1, Rc: 1, Ic: 1 &S. Kiyota (Ibaraki, Japan)\\
53823.81339	&---	        &V: 1               &H. Maehara (Saitama, Japan) \\
53824.78819	&53823.83194	&V: 1, Rc: 1, Ic: 1 &S. Kiyota (Ibaraki, Japan)\\
53827.75850	&53827.76653	&V: 1, Rc: 1, Ic: 1 &K. Nakajima (Mie, Japan)\\
53827.81163	&---	        &V: 1               &H. Maehara (Saitama, Japan)\\
53828.73766	&53828.79404	&V: 1, Rc: 1, Ic: 1 &K. Nakajima (Mie, Japan)\\
53830.81929	&53830.82405	&V: 1, Rc: 1, Ic: 1 &H. Maehara (Saitama, Japan)\\
53834.24931	&---	        &V: 1               &D. J. Mendicini (Spain)\\
53837.44860	&53837.44990	&B: 1, V: 1         &D. J. Mendicini (Spain)\\
53855.78171	&53855.78499	&V: 1, Rc: 1, Ic: 1 &H. Maehara (Saitama, Japan)\\
53857.70069	&53857.70138	&V: 1, Rc: 1, Ic: 1 &S. Kiyota (Ibaraki, Japan)\\
53859.64028	&53859.64097	&V: 1, Rc: 1, Ic: 1 &S. Kiyota (Ibaraki, Japan)\\
53886.64912	&53886.65486	&V: 1, Rc: 1, Ic: 1 &S. Kiyota (Ibaraki, Japan)\\
\hline\hline
 \end{tabular}}
\caption{Photometric observation log.}\label{tbl:photobslog}
\end{table}

\begin{table}
\centering
{\footnotesize
 \begin{tabular}{llll}
\hline\hline
\multicolumn{1}{c}{Start} & \multicolumn{1}{c}{End}&Frames &Remark\\
\multicolumn{1}{c}{(MJD)}  &\multicolumn{1}{c}{(MJD)} &\\
\hline\hline
\multicolumn{4}{l}{Organization:  VSOLJ.}\\
53824.80267	&53824.82122	&C: 24                 &K. Nakajima (Mie, Japan)\\
53825.75694	&53825.75833	&V: 1 , Rc: 1, Ic: 1    &S. Kiyota (Ibaraki,Japan)\\ 
53825.78413	&53825.82137	&V: 1 , Rc: 1, Ic: 1    &K. Nakajima (Mie, Japan)\\
53827.74097	&53827.79264	&V: 1 , Rc: 1, Ic: 98   &S. Kiyota (Ibaraki, Japan)\\
53828.71944	&53828.81230	&V: 1 , Rc: 1, Ic: 184  &S. Kiyota (Ibaraki, Japan)\\
53830.73455	&53830.79178	&V: 11, Rc: 1, Ic: 143 &S. Kiyota (Ibaraki, Japan)\\
53842.14863	&53842.17043	&Rc: 59                &H. Maehara (Saitama,Japan)\\
53850.66944	&53850.78032	&Rc: 204              &K. Nakajima (Mie,Japan)\\
53850.69490	&53850.79198	&V: 67, Rc: 65, Ic: 66 &S. Kiyota (Ibaraki,Japan)\\
\hline
\multicolumn{4}{l}{Telescope:  IRSF 1.4m~\cite{kandori06}.  Observatory:  SAAO\@.  PI: T. Nagata.}\\
53830.15679     &53830.15889    &J: 3, H: 3, Ks: 3\\
\hline\hline
 \end{tabular}}
\addtocounter{table}{-1}
\caption{Photometric observation log. (Cont'd)}
\end{table}

\begin{table}
\centering
{\footnotesize
 \begin{tabular}{lllll}
\hline\hline
\multicolumn{2}{c}{Start} & \multicolumn{2}{c}{End}& Remark\\
           &\multicolumn{1}{c}{(MJD)} &&\multicolumn{1}{c}{(MJD)}\\
\hline\hline
\multicolumn{5}{l}{Telescope: RATAN-600.  Observatory: SAO RAS\@.  PI: S. Trushkin.}\\
2006/04/01 &(53826) &2006/05/22 &(53877) &1.0 GHz--21.7 GHz\\
\hline
\multicolumn{5}{l}{Telescope: RTF-32.  Observatory:  IAA RAS\@.    PI: S. Trushkin.}\\
2006/04/08 01:06  &(53833.0458) &2006/04/08 08:52  &(53833.3694) &4 expos.  2.3 GHz 8.45 GHz\\
2006/04/09 00:53  &(53834.0368) &2006/04/09 08:50  &(53834.3680) &5 expos.\\
\hline
\multicolumn{5}{l}{Telescope: Nobeyama Millimeter Array. Observatory: Nobeyama.  PI: K. Nakanishi}\\
2006/04/25 22:10 &(53850.9236) &2006/04/25 22:30 &(53850.9375) &110.21 GHz, 98.201 GHz\\
\hline\hline
 \end{tabular}}
\caption{Radio observation log.}\label{tbl:radiobslog}
\end{table}


\begin{thebibliography}{99}
%\8.8I, \\emph{%t}, \\emph{%j} {\\bf %v} (%y) %p.
{

\bibitem{band86} D.~L.~Band, and J.~E.~Grindlay,  \emph{Synchrotron and inverse Compton 
emission from expanding sources in jets - Application to SS 433}, 
\emph{\apj} {\bf 311} (1986) 595.  

\bibitem{band89} D.~L.~Band, \emph{Comparison of Einstein and EXOSAT X-ray observations of 
SS 433}, \emph{\apj} {\bf 336} (1989) 937.  

\bibitem{chakrabarti05} S.~K.~Chakrabarti, B.~G.~Anandarao, S.~Pal, S.~Mondal, A.~Nandi, 
A.~Bhattacharyya, S.~Mandal, R.~Sagar, et al.,  \emph{SS 433: 
results of a recent multiwavelength campaign}, \emph{\mnras} {\bf 362} 
(2005) 957.  

\bibitem{cherepashchuk05} A.~M.~Cherepashchuk, R.~A.~Sunyaev, S.~N.~Fabrika, K.~A.~Postnov, 
S.~V.~Molkov, E.~A.~Barsukova, E.~A.~Antokhina, T.~R.~Irsmambetova, et al.,  \emph{INTEGRAL observations of SS433: Results of a coordinated 
campaign}, \emph{\aap} {\bf 437} (2005) 561.  

\bibitem{fabrika04} S.~Fabrika,   \emph{The jets and supercritical accretion disk in SS433}, 
\emph{Astrophysics and Space Physics Reviews} {\bf 12} (2004) 1.  


\bibitem{gies02} Gies, D.~R., M.~V.~McSwain, R.~L.~Riddle, Z.~Wang, P.~J.~Wiita, and 
D.~W.~Wingert,  \emph{The Spectral Components of SS 433}, \emph{\apj} {\bf 
566} (2002) 1069.  

\bibitem{gladyshev87} S.~A.~Gladyshev,
 				V.~P.~Goranskii, A.~M.~Cherepashchuk,
 \emph{Photometric Investigations of SS~433---Results of 1979--1986 
 Observations}, \emph{Sov.\ Astron.} {\bf 31} (1987) 541.

\bibitem{kandori06} R.~Kandori, N.~Kusakabe, M.~Tamura, Y.~Nakajima, T.~Nagayama, 
C.~Nagashima, J.~Hashimoto, J.~Hough, et al.,  \emph{SIRPOL: a 
JHK$_{\rm s}$-simultaneous imaging polarimeter for the IRSF 1.4-m telescope}, 
\emph{\procspie} {\bf 6269} (2006).

\bibitem{kato04} T.~Kato, M.~Uemura, R.~Ishioka, D.~Nogami, C.~Kunjaya, H.~Baba, and 
H.~Yamaoka,  \emph{Variable Star Network: World Center for Transient Object 
Astronomy and Variable Stars}, \emph{\pasj} {\bf 56} (2004) 1.  

\bibitem{kotani99} T.~Kotani, D.~Band, A.~M.~Cherepashchuk, R.~M.~Hjellming, N.~Kawai, 
M.~Matsuoka, M.~Namiki, T.~Oka, et al.,  \emph{Multi-wavelength 
observations of the jet sources SS 433 and XTE J1748-288}, 
\emph{Astronomische Nachrichten} {\bf 320} (1999) 335.  


\bibitem{kotani05} T.~Kotani,
 N.~Kawai,
 K.~Yanagisawa,
 J.~Watanabe,
 M.~Arimoto,
 H.~Fukushima,
 T.~Hattori,
 M.~Inata, et al., \emph{MITSuME---Multicolor Imaging Telescopes for Survey and Monstrous
Explosions}, in proceedings of \emph{4th Workshop on Gamma-Ray Burst in
	the Afterglow Era}, (2005) 755.

\bibitem{kotani06} T.~ Kotani, S.~A.~Trushkin, R.~Valiullin, K.~Kinugasa, S.~Safi-Harb, 
N.~Kawai, and M.~Namiki,  \emph{A Massive Jet Ejection Event from the 
Microquasar SS 433 Accompanying Rapid X-Ray Variability}, \emph{\apj} {\bf 
637} (2006) 486.  

\bibitem{koyama06} K.~Koyama, H.~Tsunemi, T.~Dotani, M.~W.~Bautz,
	K.~Hayashida, T.~Tsuru, H.~Matsumoto, Y.~Ogawara, et al.,
 	\emph{X-Ray Imaging Spectrometers (XIS) on Board Suzaku}, \emph{\pasj}, (2006) in press.

\bibitem{margon84} B.~Margon,   \emph{Observations of SS 433}, \emph{\araa} {\bf 22} (1984).
507.  

\bibitem{margon89}  B.~Margon, S.~F.~Anderson, \emph{Ten years of SS 433 kinematics}, 
\emph{\apj} {\bf 347} (1989) 448. 

\bibitem{mitsuda06} K.~Mitsuda, M.~Bautz, H.~Inoue, R.~Kelley,
 	K.~Koyama, H.~Kunieda, K.~Makishima, Y.~Ogawara, et al.,
 	\emph{The X-Ray Observatory Suzaku}, \emph{\pasj}, (2006) in press.

\bibitem{obayashi04} H.~Obayashi,  O.~Hashimoto, E.~Nishihara, and K.~Kinugasa,  \emph{The 
 advantages of Gunma Astronomical Observatory}, in proceedings of \emph{Third Rome Workshop on Gamma-Ray Bursts in the Afterglow Era}, (2004) 540.  


\bibitem{seaquist82} E.~R.~Seaquist, W.~S.~Gilmore, K.~J.~Johnston, and J.~E.~Grindlay,  
\emph{Simultaneous radio and X-ray activity in SS 433}, \emph{\apj} {\bf 
260} (1982) 220.  

\bibitem{trushkin04} S.~A.~Trushkin, N.~N.~Bursov, and N.~A.~Nizhelskij, 
\emph{The multifrequency monitoring of microquasars. SS433},
in \emph{Bulletin of the Special Astrophysical Observatory} {\bf 56} (2003) 57 [{\tt astro-ph/0403037}].


\bibitem{yokoo94} T.~Yokoo,  J.~Arimoto, K.~Matsumoto, A.~Takahashi, and K.~Sadakane,  
\emph{VRI CCD photometric observations of SN 1994I in M51}, \emph{\pasj} 
{\bf 46} (1994) L191.
} 
\end{thebibliography}
\end{document}